\documentclass[10pt,letterpaper,twocolumn]{article} %% two column, final layout

\usepackage{ol2}
\usepackage[draft]{hyperref}
\usepackage{amsmath}

\begin{document}

\twocolumn[ %% activate for two-column option

\title{Refractionless propagation of discretized light}

%% For REVTeX it is possible to automate superscript and e-mail callouts with the superscriptaddress option; see REVTeX4 documentation.

\author{Stefano Longhi}

\address{Dipartimento di Fisica, Politecnico di Milano and Istituto di Fotonica e Nanotecnologie del Consiglio Nazionale delle Ricerche, Piazza L. da Vinci 32, I-20133 Milano, Italy (stefano.longhi@polimi.it)}

\begin{abstract}
Light refraction, i.e. the bending of the path of a light wave at the interface between two different dielectric media, is ubiquitous in optics. Refraction arises from the different speed of light and is unavoidable in continuous media according to Snell's Law. Here we show rather counterintuitively that omnidirectional refractionless propagation can be observed for discretized light crossing a tilted interface separating two homogeneous waveguide lattices.
\end{abstract}

 ] %% activate for two-column option

Wave refraction, i.e. the bending of the path of a wave as it propagates in an inhomogeneous medium or at the interface between two different media,  is ubiquitous in wave physics. Refraction arises from a change in the speed of the wave and is commonly observed for classical and quantum waves. In optics, refraction is caused by a change of the refractive index and is described quantitatively by Snell's Law. A sharp change of the refractive index is also responsible for wave reflection. While several methods are known to avoid wave reflection, such as the use of antireflection coatings \cite{r4,r5}, adiabatic index matching \cite{r6,r7,r8,r9} or specially-tailored index profiles \cite{r10,r11,r12,r13,r14,r15}, wave refraction seems unavoidable because of refractive index change. Material structuring can change light refraction, for example in negative-index media \cite{r16,r17} or in photonic crystals \cite{r18,r19} light can be forced to bend in the wrong way (negative refraction). However, omnidirectional refractionless propagation of light waves at a sharp or smooth interface separating two different continuous media is unlikely.\par Wave refraction and reflection are not restricted to light propagation in continuous media; they are observed also when light behavior is discretized \cite{r20}, i.e. when transport arises from evanescent mode coupling in guiding structures. Discrete light propagation plays a crucial role in integrated classical and quantum photonics, where the flow of light can be manipulated in
unprecedented ways \cite{r20,r21,r22,r23}. Like for continuous media, Snell's Law for reflection and refraction of light in discrete optical media can be derived \cite{r24}, and effects such as anomalous or negative refraction can be observed \cite{r25,r26,r26bis}. Reflection can be suppressed for both continuous and discrete light propagation. However, in a recent work \cite{r27} it has been shown that reflectionless propagation across potential barriers or defects can be realized in discrete optics under much less restricting conditions than those usually required in continuous media. A natural question than arises: can we cancel light refraction in discrete optics?\\ 
In this Letter we show that, contrary to the common wisdom that considers refraction unavoidable between media with different refractive indices, omnidirectional refractionless propagation can be realized for discretized light {\em at any arbitrarily-shaped} and suitably tilted interface separating two homogeneous waveguide lattices with different effective indices. The refractionless effect is observed for a sufficiently high effective index mismatch and for an interface tilted at a suitable angle above the light cone of the lattice band. Refractionless propagation is possible owing to the discrete (rather than continuous) translational invariance of the lattice and disappears in the continuous limit.\\
  To highlight the major role of discrete versus continuous translational invariance  in determining refractionless light propagation, let us first recall the phenomenon of light refraction at a sharp or smooth interface separating two continuous dielectric media with different refractive indices $n_1$ and $n_2$ [Fig.1(a)]. The refractive index $n=n(X)$ varies along the $X$ direction normal to the interface, with $n(X) \rightarrow n_{1,2}$ as $ X \rightarrow \pm \infty$. We refer space to a coordinate system $(x,y,z)$ such that the $X$ axis lies in  the $(x,z)$ plane and is tilted with respect to the $x$ axis, namely $X=x+vz$ with $v >0$; see Fig.1(a). Note that, for $v \ll 1$, $v$ is the tilt angle of the interface with respect to the $z$ axis. Clearly the problem of wave reflection and refraction at the interface does not depend on the tilt $v$ of the interface, since a non vanishing value of $v$ just corresponds to a shift of  the angles of incident, reflected and transmitted waves from the reference $z$ axis. However, as we will show below this is not the case of refraction and reflection for discretized light. Assuming TE-polarized waves at frequency $\omega$, with the electric $\mathbf{E}$-field parallel to the $y$ axis, $\mathbf{E}=E_y(x,z) \mathbf{u}_y$, the $E_y$ amplitude satisfies the Helmholtz equation
  \begin{equation}
  \frac{\partial^2 E_y}{\partial z^2}+\frac{\partial^2 E_y}{\partial x^2}+k_0^2 n^2(x+vz) E_y=0
  \end{equation}
  where $k_0= \omega/c_0$ is the wave number in vacuum. For left-incidence side and above the critical angle (to avoid total internal reflection in case $n_2<n_1$), an optical beam propagating at an angle $\theta_i$ with the vertical $z$ axis [Fig.1(a)] is refracted to the angle $\theta_t$ given by the Snell's Law
    \begin{equation}
  n_1 \cos(\theta_i+v)=n_2 \cos (\theta_t+v).
  \end{equation}
  For an adiabatic interface, wave reflection can be neglected and the beam is fully transmitted, whereas reflection is observed at a sharp interface when the refractive index $n(X)$ changes on the spatial scale of the optical wavelength. To compare the different behavior of refraction in continuous versus discrete media,
let us consider the limiting case of grazing incidence, a small tilt angle $v$  and a small refractive index change, i.e. $n(X)=n_1+ \Delta n(X)$ with $|\Delta n(X)| \ll n_1$, which implies $n_{2} \simeq n_1$.  After setting $E_y(x,z)= \psi(x,z) \exp(ik_0 n_1 z)$ with $\psi(x,z)$ slowly-varying with $z$ over the spatial scale of the optical wavelength $ \sim 1 / k_0$, from Eq.(1) the paraxial optical wave equation is obtained
  \begin{equation}
   i \hbar \frac{\partial \psi}{\partial z}=- \frac{\hbar^2}{2m} \frac{\partial^2 \psi}{\partial x^2}+V(x+vz) \psi
  \end{equation}
    where we have set $\hbar \equiv 1/ k_0$, $m=n_1 \simeq n_2$ and where $V(x+vz) \equiv [n_1^2-n^2(x+vz)]/(2 n_1) \simeq n_1-n(x+vz)$ is the so-called optical potential. The paraxial wave equation (3) is formally analogous to the non-relativistic Schr\"odinger equation that describes one-dimensional scattering (reflection and transmission) of matter waves from a potential step $V$, provided that the spatial coordinate $z$ is replaced with time \cite{r28}. Note that in the quantum mechanical problem a non vanishing value of interface tilt $v$ corresponds to a drift of the potential step with a drift velocity $v$. The incidence and refracted angles $\theta_i$ and $\theta_t$ are related to the particle momentum in the $x+vz \rightarrow -\infty$ and $x+vz \rightarrow \infty$ spatial regions, and can be obtained from the Snell's Law (2) in the small angle limit.
 % \begin{equation}
 % \theta_t \simeq -v+ \sqrt{2 \left( 1-\frac{n_1}{n_2} \right) +\frac{n_1}{n_2} \left( \theta_i+v \right)^2}.
  % \end{equation}
Quantum mechanically, the Snell's Law of refraction corresponds to conservation of the total particle energy after being transmitted across the potential step in a moving reference frame where the potential appears at rest. The fact that wave refraction does not depend on the tilt $v$ of the interface, i.e. on the drift velocity of the potential step, is basically related to the Galilean invariance of the non-relativistic Schr\"odinger equation \cite{r29}. In the moving reference frame 
\begin{equation}
 X=x+vz \; , \; Z=z,
\end{equation}
where the potential is at rest,  the Schr\"odinger equation (3) acquires an additional drift term $ \sim v ({\partial \psi}/{\partial X})$, which however can be removed after a gauge transformation. This is possible  because of the parabolic shape of the energy-momentum dispersion relation of plane waves for non-relativistic particles.\\ 
A completely different scenario is found when considering refraction of discretized light in a waveguide lattice with a potential step [Fig.1(b)], which is described by a discrete version of the  Schr\"odinger equation. As discussed in recent works \cite{r27,r30}, the discrete Schr\"odinger equation with a sinusoidal (rather than parabolic) energy dispersion relation is not invariant under a Galilean transformation. Therefore the refraction properties of discretized light are expected to be deeply modified by the  drift velocity $v$, i.e. by the interface tilt.\\ 
We consider a linear array of waveguides equally spaced by the distance $a$ along the $x$-axis. The propagation constant $\beta$ (i.e. effective mode index) of waveguides is assumed to vary across the array to describe a tilted potential step, taking the two asymptotic values $\beta_1$ and $\beta_2$ as $ X \rightarrow \pm \infty$ [Fig.1(b)]. Let us indicate by $c_n(z)$ the modal amplitude of light trapped in the $n-th$ waveguide of the array ($n=0, \pm 1, \pm 2,...$) at the propagation distance $z$, and let $\psi(x,z)$ be a complex amplitude of continuous variables $x$ and $z$ such that $c_n(z)=\psi(x=na,z)$. In the nearest-neighbor and tight-binding approximations, the evolution equation of the amplitude $\psi(x,z)$  is described by the discrete Schr\"odinger equation  \cite{r20,r21,r22,r27,r28}
\begin{equation}
i \frac{\partial \psi}{\partial z}=- \kappa [\psi(x+a,z)+\psi(x-a,z)] +\beta(x+vz) \psi(x,z)
\end{equation}
 \begin{figure}[htb]
\centerline{\includegraphics[width=8.4cm]{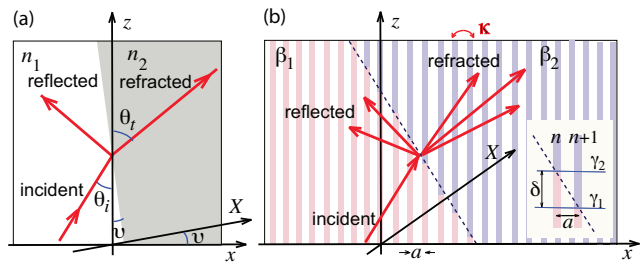}} \caption{ \small
(Color online) Schematic of optical refraction and reflection at a tilted refractive index step (a) in continuous dielectric media, and (b) in waveguide arrays (discretized light). Multiple refraction and reflection can occur in the latter case. In (a) refraction is unavoidable unless $n_2=n_1$, whereas in (b) omnidirectional refractionless propagation is possible for $\beta_2 \neq \beta_1$. The inset in (b) shows that light waves in adjacent waveguides $n$ and $(n+1)$ acquire a longitudinal phase difference $\Delta \phi=(\beta_2-\beta_1)\delta$ while crossing the interface, from plane $\gamma_1$ to plane $\gamma_2$, where $\delta=a/v$. Refractionless propagation is found when $\Delta \phi$ is an integer multiple than $ 2 \pi$.}
\end{figure} 
where $\kappa$ is the coupling constant between adjacent waveguides and $\beta=\beta(x+vz)$ is the  propagation constant, which describes a tilted potential step (either sharp or smooth); see Fig.1(b). For a straight interface ($v=0$), discrete refraction and the discrete version of the Snell's Law, previously studied in Ref.\cite{r24}, can be readily determined by considering the lattice dispersion bands for Bloch waves $\psi(x,z) \sim \exp[iqx-i E(q)z]$ in the two asymptotic and spatially-homogenous regions $x \rightarrow \pm \infty$ of the array, which are given by $E(q)=E_1(q)=- 2 \kappa \cos (qa)+\beta_1$ for $ x \rightarrow -\infty$ and $E(q)=E_2(q)=- 2 \kappa \cos (qa)+\beta_2$ for $ x \rightarrow \infty$. The two dispersion curves are the same but displaced by the amount $\beta_2-\beta_1$. To observe discrete refraction in the $v=0$ case, the two bands should be partially overlapped, i.e. the constraint $|\beta_2-\beta_1|< 4 \kappa$ should be satisfied, otherwise the discrete analogue of total internal reflection (Bragg scattering) is observed for any incident wave packet.  For left incidence side, we consider an incident wave packet with carrier Bloch wave number $q_1$ ($0<q_1 < \pi/a$), which propagates transversely along the array with the group velocity $v_{g1}=(\partial E_1/ \partial q)_{q=q_1}=2 \kappa a \sin (q_1a )>0$. The refracted (transmitted) wave packet has a shifted carrier Bloch wave number $q_2$ which is determined from propagation constant (energy) conservation rule $E_2(q_2)=E_1(q_1)$, i.e.
  \begin{figure}[htb]
\centerline{\includegraphics[width=8.7cm]{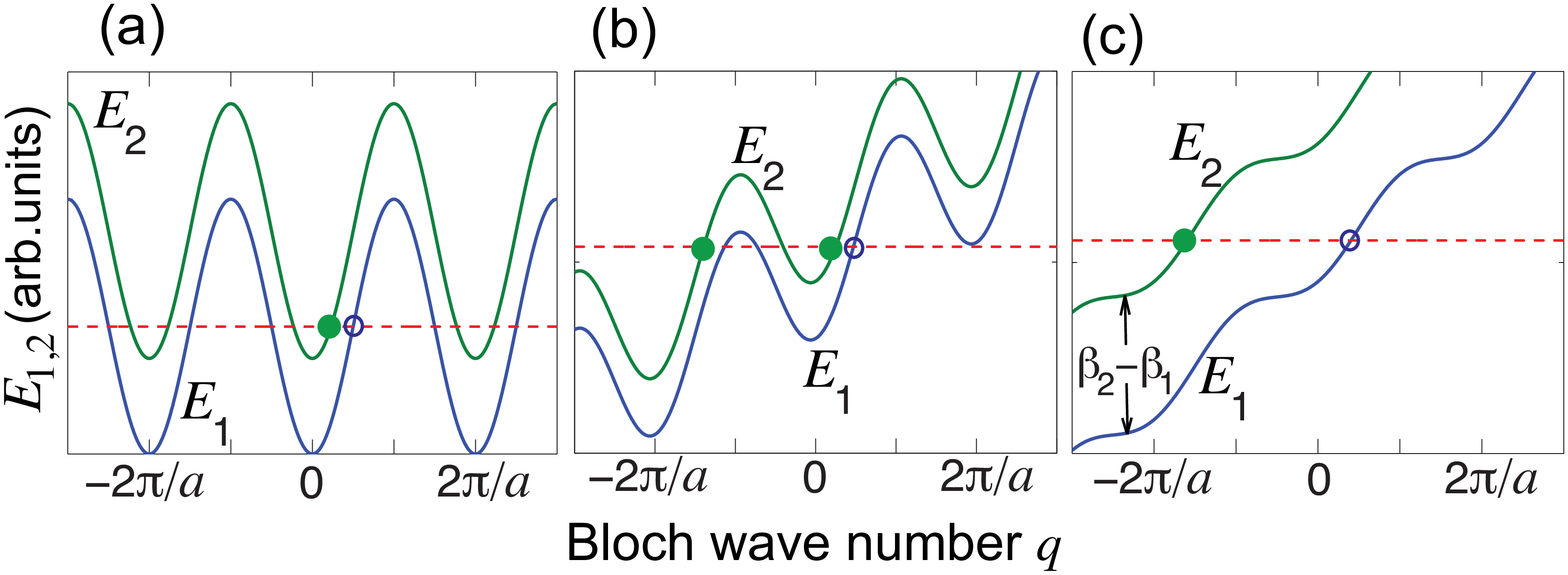}} \caption{ \small
(Color online) Lattice dispersion curves $E_{1}(q)$ and $E_2(q)=E_1(q)+(\beta_2-\beta_1)$ (solid lines) of the two waveguide arrays far from the interface in the tilted reference frame $(X,Z)$ for (a) $v=0$, (b) $v<v_c$, and (c) $v> v_c$, where $v_c=2 \kappa a$ is the largest velocity of propagative waves allowed by the light cone of the lattice band. The open circle corresponds to the Bloch wave number $q_1$ of the incident wave [$q_1= \pi/(2a)$ in the figure]. The Bloch wave numbers $q_2$ of transmitted waves, obtained from the conservation law $E_1(q_1)=E_2(q_2)$, are indicated by filled circles. In (a) and (c) there is only one allowed value of $q_2$, whereas in (b) two values of $q_2$, with different group velocities, are possible (double refraction). In (a) multiple roots $q_2$ are disregarded since they differ each other by integer multiplies than $ 2 \pi /a$, and thus correspond to the same Bloch wave. }
\end{figure} 
\begin{equation}
\cos (q_2 a)=\cos(q_1 a)+(\beta_2-\beta_1)/(2 \kappa).
\end{equation}
with the constraint $\sin(q_2 a)>0$ [Fig.2(a)]. The transmitted wave packet propagates with the transverse group velocity $v_{g2}=(\partial E_2/ \partial q)_{q=q_2}=2 a \kappa \sin (q_2 a)$, which is generally different than $v_{g1}$, yielding a bending of the transmitted beam (discrete refraction \cite{r24}). Besides refraction, a reflected wave packet is  observed for a sharp potential step. An example of discrete refraction for $v=0$ at a sharp potential step [$\beta(X)=\beta_1$ for $X<0$ and $\beta(X)=\beta_2$ for $X>0$] is shown in Fig.3(a). The figure depicts the numerically-computed evolution, along the spatial propagation distance $z$,  of a broad and Gaussian-shaped discretized beam with carrier Bloch wave number $ q_1= \pi/(2a)$. Clearly, the refracted beam is bent after crossing the potential step according to the discrete version of the Snell's Law. The refraction/reflection problem of the continuous Schr\"odinger equation discussed above is obtained from the discrete Schr\"odinger equation (5) by assuming that $\psi(x,z)$ and $\beta(x,z)$ vary slowly with $x$ over the spatial scale of the lattice period $a$ and provided that $|\beta_2-\beta_1| \ll 2 \kappa$. In this limiting case the difference $\psi(x+a)+\psi(x-a)$ can be approximated by the second-order derivative $\psi(x+a)+\psi(x-a) \simeq 2 \psi(x,z)+a^2 (\partial^2 \psi/ \partial x^2)$, and Eq.(5) takes the form of the continuous Schr\"odinger equation (3). Note that in such a limiting case the discrete translational invariance of the lattice and associated Bragg reflection at the Brillouin zone edges are lost since the sinusoidal dispersion relation of the lattice band is approximated by the parabolic dispersion curve near the bottom of the band. Therefore, to observe refractionless propagation in the lattice we should operate in a regime where the discrete translational invariance of the lattice is kept and refraction is sensitive to the transverse tilt $v$ of the interface potential. We are now going to prove the following property: provided that the condition
\begin{equation}
|\beta_2-\beta_1| \geq 4 \pi \kappa
\end{equation}
is met,  i.e. for a sufficiently high potential step, omnidirectional refractionless propagation of discretized light across an arbitrarily shaped interface is observed for a tilt $v$ satisfying the condition
\begin{equation}
v=|\beta_2-\beta_1|a/(2 \pi).
\end{equation} 
Such a general result can be demonstrated by considering the refraction problem for the discrete Schr\"odinger equation in the tilted reference frame $(X,Z)$ defined by Eq.(4), where the potential step becomes $Z$-invariant and Eq.(5) takes the form
\begin{equation}
i \frac{\partial \psi}{\partial Z}=-i v \frac{\partial \psi}{\partial X}- \kappa [\psi(X+a,Z)+\psi(X-a,Z)]+ \beta(X) \psi(X,Z).
\end{equation}
Contrary to the continuous Schr\"odinger equation, the drift term $ \sim v (\partial \psi / \partial X)$ on the right hand side of Eq.(9) can not be removed by a gauge transformation \cite{r27,r30}, and like for reflection \cite{r27} we expect the refraction problem to be sensitive to the tilt $v$. In the $(X,Z)$ reference frame, 
a discretized optical wave with carrier wave number $q_1$ on the left side of the interface has an effective propagation constant $E=E_1(q=q_1)$, where $E_1(q)=- 2 \kappa \cos(qa)+vq+ \beta_1$ is the lattice dispersion curve for the homogeneous array at  the left side of the interface \cite{r27}. The group velocity of the beam is given by $v_{g1}=(\partial E_1/ \partial q)_{q=q_1}=2 \kappa a \sin (q_1 a)+v$ with $v_{g1}>0$. Owing to conservation of the effective propagation constant, the transmitted (refracted) propagative wave on the right hand side of the interface has a shifted Bloch wave number $q_2$ satisfying the condition $E_2(q_2)=E_1(q_1)$, where  $E_2(q)=- 2 \kappa \cos(qa)+vq+ \beta_2$  is the lattice dispersion curve  for the homogeneous array at  the right side of the interface. Hence $q_2$ is found as the real-valued root of the equation
\begin{equation}
-2 \kappa \cos(q_2 a)+q_2v+\beta_2=-2 \kappa \cos(q_1 a)+q_1 v+ \beta_1.
\end{equation}
The group velocity of the refracted wave is given by $v_{g2}=(\partial E_2/ \partial q)_{q=q_2}=2 \kappa a \sin (q_2 a)+v$, and hence the acceptable solutions to Eq.(10) are those with $v_{g2}>0$. The real-valued roots of the  transcendental equation (10) can be readily determined by the geometric construction shown in Fig.2. Clearly, for small values of $v$, there exist more than one solutions $q_2$ to Eq.(10) with different and positive group velocities $v_{g2}$ [Fig.2(b)]: this means that the refracted wave breaks into two (or more) wave packets, i.e. double (or multiple) refraction is found. Such a result is confirmed by direct numerical simulations of coupled-mode equations in the waveguide lattice with a sharp and slightly-tilted potential step [Fig.3(b)]. Note that multiple refraction arises here from interfacing two different waveguide lattices in the single-band approximation, and therefore it is rather distinct than beam breakup and multiple refraction observed in a homogeneous lattice (i.e. without any interface) when several lattice bands are excited  \cite{rreferee}. As the tilt $v$ is increased above the critical value $v_c= 2 \kappa a$, i.e. above the light cone defined by the lattice band dispersion relation \cite{r27}, a single solution to Eq.(10) with positive group velocity does exist, i.e. double (or multiple) refraction is prevented. In this regime also reflection is prevented \cite{r27}. Omnidirectional refractionless propagation is found provided that $v_{g2}=v_{g1}$ for an arbitrary value of the  Bloch wave number $q_1$ of the incoming wave. This requirement is readily satisfied whenever $q_2$ differs from $q_1$ by integer multiplies than $ 2 \pi/a$. Using Eq.(10) with $q_2=q_1\pm 2 \pi N/a$ one then obtains the simple condition
$v=|\beta_2-\beta_1| a/(2 \pi  N)$ for the tilt, where $N$ is a positive integer. The largest tilt $v$ is obtained by taking $N=1$, which thus yields Eq.(8). To avoid the appearance of other refracted wave packets, we should require $v \geq v_c$, which provides a lower limit for the height of the potential step given by Eq.(7). We remark that the refractionless property, stated by Eqs.(7) and (8), is valid for an {\em arbitrary} shape (sharp or smooth) of the potential step $\beta(X)$. A simple physical explanation of the condition (8) can be gained in the limiting case of a sharp potential step by considering the phase delay between light waves in adjacent waveguides when they cross  
the interface, as illustrated in the inset of Fig.1(b). Owing to the difference in the propagation constants $\beta_1$ and $\beta_2$, the light waves in the two waveguides crossing the interface acquire, from plane $\gamma_1$ to plane $\gamma_2$, different optical phases, with a resulting phase delay $\Delta \phi= (\beta_2-\beta_1) \delta$, where $\delta=a/v$ is the distance between planes $\gamma_1$ and $\gamma_2$. Such a phase delay is basically equivalent to a change of the Bloch wave number $q$ (from $q_1$ to $q_2$), which is responsible for refraction. However, provided that $\Delta \phi$ is a multiple of $ 2 \pi$, i.e. $\Delta \phi= 2 \pi N$,  refraction is cancelled. For $N=1$, one obtains Eq.(8). The constraint imposed by Eq.(7) can be qualitatively understood by observing that the above explanation holds provided that coupling between adjacent waveguides remains negligible over the propagation distance $\delta$, i.e. $\delta=a/v$ should be much smaller than the coupling length $ \sim \pi/ \kappa$. Since $v=|\beta_2-\beta_1|a/(2 \pi)$, negligible coupling requires  $|\beta_2-\beta_1| \gg 2 \kappa$.\\
We checked the occurrence of refractionless propagation in a waveguide lattice with a tilted refractive index step by numerical simulations of coupled-mode equations. Figure 4, left panels, shows an example of refractionless propagation of a discretized Gaussian beam across a tilted interface for a few increasing values of the incidence angle, i.e. Bloch wave number $q_1$. A sharp step-index profile, satisfying the constraint (7), is assumed ($\beta_1-\beta_2=4 \pi \kappa$), and the tilt angle $v$ of the interface is set according to Eq.(8) ($v=2  a \kappa$). Clearly, regardless of the incidence angle, the transmitted beam is not bent, i.e. omnidirectional refractionless propagation across the index step is observed. For comparison, the right panels in Fig.4 show the numerically-computed beam propagation for the 'wrong' tilt $v=2.4 \kappa a$: while the tilt $v$ is larger than the critical value $v_c=2 \kappa a$ and thus multiple refraction is avoided, the refracted beam propagates at a different inclination angle, i.e. refraction is not suppressed. Finally, it should be noted that, if condition (8) for the tilt is satisfied but Eq.(7) is not, one generally observes multiple refraction, with only one transmitted beam being not deviated; see Fig.3(b).

   \begin{figure}[htb]
\centerline{\includegraphics[width=9cm]{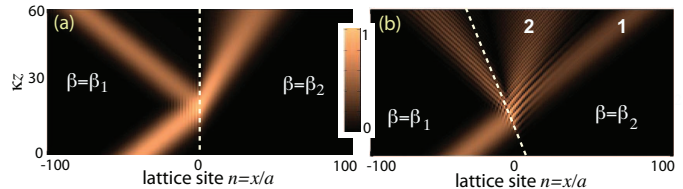}} \caption{ \small
(Color online) Refraction of a discretized Gaussian beam from a sharp refractive index step $\beta$  on a waveguide lattice for (a) straight and (b) tilted interfaces. Parameter values are 
$\beta_{1}-\beta_2=1.8 \kappa$, $v=0$ in (a), and  $\beta_{1}-\beta_2=2.6 \kappa$, $v=|\beta_2-\beta_1|a/(2 \pi) \simeq 0.4138 \kappa a$ in (b). The panels show on a pseudo color map the evolution of the waveguide field amplitudes $|c_n(z)|$ versus longitudinal propagation distance $z$ as obtained by numerical simulations of coupled-mode equations with the initial condition $c_n(0) \sim \exp\{-[(n+40)/15]^2+i n q_1a \}$ and $q_1a= \pi/2$. The bold dashed lines depict the interface of the sharp index step ( $\beta=\beta_1$ and $\beta=\beta_2$ on the left and right sides of the lines, respectively).  In case of a straight interface [panel (a)] there is only one refracted and one reflected beam.  In (b) (tilted interface) the condition (8) is satisfied, however since the tilt $v$ is smaller than the critical value $v_c=2 \kappa a$ double refraction is found: while refracted beam 1 is not bent, the other refracted beam 2 is bent.}
\end{figure}  

  \begin{figure}[htb]
\centerline{\includegraphics[width=9cm]{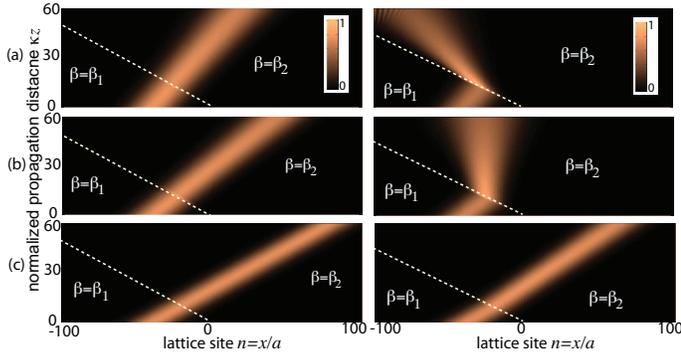}} \caption{ \small
(Color online) Left panels: Refractionless propagation of a discretized Gaussian beam across a sharp refractive index step $\beta$ on a waveguide lattice for parameter values  $\beta_{1}-\beta_2=4 \pi \kappa$,  $v=2 \kappa a$ and for a few increasing values of the Bloch wave number $q_1$ of the incident wave: (a) $q_1= \pi/(6a)$, (b) $q_1= \pi/(4a)$, and (c) $q_1=\pi/(2a)$. The panels depict on a pseudo color map the evolution of the field amplitudes $|c_n(z)|$ versus longitudinal distance $z$ for an initial Gaussian-shaped beam distribution $c_n(0) \sim \exp\{-[(n+40)/15]^2+i n q_1a \}$. Right panels: same as left panels, but for a different tilt of the interface ($v=2.4 \kappa a$). Note that, since in both left and right panels the tilt $v$ is equal or larger than the critical drift $v_c= 2 \kappa a$, there is not any reflection \cite{r27}.}
\end{figure}  

 In conclusion, refraction is a universal phenomenon observed when light crosses the boundary of dielectric media with different optical density. It is at the heart of important  effects, such as light focusing, lensing, guiding and bending. In effectively continuous media, refraction can be engineered by material structuring, and can be even reversed like in negative-index metamaterials. However, a common belief is that refraction is unavoidable unless equal effective indices are realized. Here we have shown that, contrary to such a common wisdom, omnidirectional refractionless propagation can be achieved when light behavior is discretized \cite{r20}. Our results shed new light into an old phenomenon of optics for a form of light transport that is becoming of great relevance in integrated classical and quantum photonics \cite{r20,r21,r22,r23}.\\
 \\
 The author acknowledges hospitality at the IFISC (CSIC-UIB).

\newpage

%%%%%%%%%%%%%%%%%%%%%%%%%%%%%%%
% References with full titles %
%%%%%%%%%%%%%%%%%%%%%%%%%%%%%%%

%\footnotesize
 {\bf References with full titles}\\
 \\
 \noindent
%1. G.F. Fitzgerald, {\it On the Electromagnetic Theory of the Reflection and Refraction of Light },  Philos. Trans. Royal Soc. London
%{\bf 171},  691 (1880).\\
%2. H.A. Lorentz, {\it On the Theory of the Reflection and Refraction of Light} (Rodopi B.V., Amsterdam-Atlanta1997).\\
%3. J. Lekner, \textit{Theory of Reflection of Electromagnetic and Partilce Waves} (Kluwer, Dordrecht, 1987).\\
1. H.A. Macleod, {\it Thin-Film Optical Filters} (Taylor \& Francis, 2009).\\
2. H.K. Raut, V.A. Ganesh, A.S. Nair, and S. Ramakrishna, {\it Anti-reflective coatings: a critical, in-depth review}, Energy Environ. Sci.{\bf 4} , 3779 (2011).\\
3. W. H. Southwell, {\it Gradient-index antireflection coatings }, Opt. Lett. {\bf 8}, 584 (1983).\\
4. J.-Q. Xi, M. F. Schubert, J. K. Kim, E. F. Schubert, M. Chen, S.-Y. Lin, W. Liu, and J. A. Smart, {\it Optical thin-film materials with low refractive index for broadband elimination of Fresnel reflection} Nat. Photonics {\bf 1}, 176 (2007).\\
5. D.J. Poxson, M.F. Schubert, F.W. Mont, E.F. Schubert, and J.K. Kim, {\it Broadband omnidirectional antireflection coatings optimized by genetic algorithm}, Opt. Lett. {\bf 34}, 728 (2009).\\
6. C.-H. Chang, J.A. Dominguez-Caballero, H.J. Choi, and G. Barbastathis, {\it Nanostructured gradient-index antireflection diffractive optics}, Opt. Lett. {\bf 36}, 2354 (2011).\\
7. I. Kay and H. E. Moses, {\it Reflectionless transmission through dielectrics and scattering potentials}, J. Appl.  Phys. {\bf 27}, 1503 (1956).\\
8. K.-H. Kim and Q.-H. Park, {\it Perfect anti-reflection from first principles}, Sci. Rep. {\bf }3, 1062 (2013).\\ 
9. L.V. Thekkekara, V.G. Achanta, and S. Dutta Gupta, {\it Optical reflectionless potentials for broadband, omnidirectional antireflection}, Opt. Express {\bf 22}, 17382 (2014).\\
10. S.A.R. Horsley, M. Artoni, and  G.C. La Rocca, {\it Spatial Kramers-Kronig relations and the reflection of waves}, Nature Photon.{\bf 9}, 436 (2015).\\
11. S. Longhi, {\it Wave reflection in dielectric media obeying spatial Kramers-Kr\"{o}nig relations}, EPL {\bf 112}, 64001 (2015).\\
12. S.A.R. Horsley, C.G. King, and T.G. Philbin, {\it Wave propagation in complex coordinates}, J. Opt. {\bf 18}, 044016 (2016).\\
13. V. G. Veselago, {\it The electrodynamics of substances with simultaneously negative values of $\epsilon$ and $\mu$}, Sov. Phys. Usp. {\bf 10}, 509 (1968).\\
14. J.B. Pendry, {\it Negative refraction}, Contemp. Phys. {\bf 45}, 191 (2004).\\
15. M. Notomi, {\it Negative refraction in photonic crystals}, Opt. Quantum Electron. {\bf 34}, 133 (2002).\\
16. E. Cubukcu, K. Aydin, E. Ozbay, S. Foteinopoulou, and  C.M. Soukoulis, {\it Electromagnetic waves: Negative refraction by photonic crystals}, Nature {\bf 423}, 604 (2003).\\
17. D. N. Christodoulides, F. Lederer, and Y. Silberberg, {\it Discretizing light behaviour in linear and nonlinear waveguide lattices}, Nature {\bf 424}, 817 (2003).\\
18. A. Szameit and S. Nolte, {\it Discrete optics in femtosecond-laser-written photonic structures}, J. Phys. B {\bf 43}, 163001 (2010).\\
19.  I.L. Garanovich, S. Longhi, A.A. Sukhorukov, and Y.S. Kivshar, {\it Light propagation and localization in modulated photonic lattices and waveguides}, Phys. Rep. {\bf 518}, 1 (2012).\\
20. T. Meany, M. Gr\"afe, R. Heilmann, A. Perez-Leija, S. Gross, M.J. Steel, M.J. Withford, and A. Szameit, {\it Laser written circuits for quantum photonics}, Laser \& Photon. Rev. {\bf 9}, 363 (2015).\\
21. A. Szameit, H. Trompeter, M. Heinrich, F. Dreisow, U. Peschel,
T. Pertsch, S. Nolte, F. Lederer, and A. T{\"u}nnermann, {\it Fresnel laws in discrete optical media}, New J. Phys. {\bf 10}, 103020 (2008).\\
22. T. Pertsch, T. Zentgraf, U. Peschel, A. Br\"auer, and F. Lederer, {\it Anomalous refraction and diffraction in discrete optical systems}, Phys. Rev. Lett. {\bf 88}, 093901
(2002).\\
23. R. Khomeriki and L. Tkeshelashvili, {\it Negative refraction and spatial echo in optical waveguide arrays}, Opt. Lett. {\bf 37}, 4419 (2012).\\
24. L. Xu, Y. Yin, F. Bo, J. Xu, and G. Zhang, {\it Anomalous refraction in disordered one-dimensional
photonic lattices}, J. Opt. Soc. Am. B {\bf 31}, 105 (2014).\\
25. S. Longhi, {\it Reflectionless and invisible potentials in photonic lattices}, Opt. Lett. {\bf 42}, 3229 (2017).\\
26. S. Longhi, {\it Quantum-optical analogies using photonic structures},  Laser \& Photon. Rev. {\bf 3}, 243 (2009).\\
27. G. Rosen, {\it Galilean Invariance and the General Covariance of Nonrelativistic Laws}, Am. J. Phys. {\bf 40}, 683 (1972).\\
28. S. Longhi, {\it Kramers-Kronig potentials for the discrete Schr\"odinger equation}, Phys. Rev. A {\bf 96}, 042106 (2017).\\
29.  D. Mandelik, H. S. Eisenberg, Y. Silberberg, R. Morandotti, and J. S. Aitchison, {\it Band-Gap Structure of Waveguide Arrays and Excitation of Floquet-Bloch Solitons}, Phys. Rev. Lett. {\bf 90}, 053902 (2003).

\end{document}